# Interactive Launch of 16,000 Microsoft Windows Instances on a Supercomputer

Michael Jones, Jeremy Kepner, Bradley Orchard, Albert Reuther, William Arcand,
David Bestor, Bill Bergeron, Chansup Byun, Vijay Gadepally, Michael Houle,
Matthew Hubbell, Anna Klein, Lauren Milechin, Julia Mullen,
Andrew Prout, Antonio Rosa, Siddharth Samsi, Charles Yee, Peter Michaleas
MIT Lincoln Laboratory, Lexington, MA, U.S.A.

*Abstract*—Simulation, machine learning, and data analysis require a wide range of software which can be dependent upon specific operating systems, such as Microsoft Windows. Running this software interactively on massively parallel supercomputers can present many challenges. Traditional methods of scaling Microsoft Windows applications to run on thousands of processors have typically relied on heavyweight virtual machines that can be inefficient and slow to launch on modern manycore processors. This paper describes a unique approach using the Lincoln Laboratory LLMapReduce technology in combination with the Wine Windows compatibility layer to rapidly and simultaneously launch and run Microsoft Windows applications on thousands of cores on a supercomputer. Specifically, this work demonstrates launching 16,000 Microsoft Windows applications in 5 minutes running on 16,000 processor cores. This capability significantly broadens the range of applications that can be run at large scale on a supercomputer.

*Keywords*—High Performance Computing, Manycore, Microsoft Windows, Wine, Windows Emulation, Knight's Landing

## I. INTRODUCTION

With the slowing down of Moore's Law [1], [2], parallel processing has become a primary technique for increasing application performance. Physical simulation, machine learning, and data analysis are rapidly growing applications that are utilizing parallel processing to achieve their performance goals. These applications require a wide range of software which can be dependent upon specific operating systems, such as Microsoft Windows. Parallel computing directly in the Windows platform has a long history [3]–[7]. The largest supercomputers currently available almost exclusively run the Linux operating system [8], [9]. Using these Linux powered supercomputers, it is possible to rapidly launch interactive applications on thousands of processors in a matter of seconds [10].

A common way to launch multiple Microsoft Windows applications on Linux computers is to use virtual machines (VMs) [11]. Windows VMs replicate the complete operating system and its virtual memory environment for each instance of the Windows application that is running, which imposes a great deal of overhead on the applications. Launching many Windows VMs on a large supercomputer can often take several seconds per VM [12]–[14]. While this performance is adequate for interactive applications that may require a handful of processors, scaling up such applications to the thousands of processors typically found in a modern supercomputer is prohibitive.

This paper describes a unique approach using the Lincoln Laboratory LLMapReduce technology in combination with the Wine Windows compatibility layer to rapidly launch and run Microsoft Windows applications on thousands of cores on a supercomputer. Specifically, this work demonstrates launching 16,000 Microsoft Windows applications in 5 minutes running on 16,000 processor cores. This capability significantly broadens the range of applications that can be run at large scale on a supercomputer.

The primary goal of this paper is to illustrate the feasibility and provide a set of baseline measurements showing the kind of performance gain that can be realized by scaling a Windows application in an pleasingly parallel manner on a modern supercomputer. The organization of the rest of this paper is as follows. Section II goes into more detail on the various technologies for running Windows in a Linux environment. Section III describes the LLMapReduce technology used to launch thousands of simultaneous Windows instances. Section IV provides details on the hardware and software environment used to perform the launch time measurements. Section V presents the performance results and an overview of the findings. Section VI summarizes the work, the benefits gained by this approach, and describes future directions.

## II. VIRTUALIZATION, CONTAINERIZATION, AND WINE

As of November 2017, 100% of the world's Top 500 supercomputers are running on the Linux operating system. [15] As a result, the vast majority of the modeling and simulation codes commonly used in science and engineering either natively run in Linux, or have a dedicated compute server component that can be separately deployed on a large number of Linux computers to render tractable the enormous computational complexity of modern models and simulations.

While the commoditization of the x86 platform and it's rapidly expanding hardware capabilities have led to expo-

This material is based upon work supported by the Assistant Secretary of Defense for Research and Engineering under Air Force Contract No. FA8721-05-C-0002 and/or FA8702-15-D-0001. Any opinions, findings, conclusions or recommendations expressed in this material are those of the author(s) and do not necessarily reflect the views of the Assistant Secretary of Defense for Research and Engineering.

nential growth in the use of virtualization as a means of maximizing the efficient use of system resources and allowing for different operating systems to coexist on a single physical machine, the overhead of full hardware virtualization - running an entire completely separate operating system kernel and full complement of system libraries subordinate to a hypervisor - is significant. In addition to these resource overhead concerns, a recent study measured the launch times of a stripped down Ubuntu Linux image on three of the most popular virtual machine provisioning systems and found that various overheads could account for up to 120 seconds of additional processing time on a modern hardware platform. [14]

Operating system-level virtualization methods such as chroot [16], FreeBSD jail [17], OpenVZ [18], User Mode Linux [19], and Linux Containers as made popular by Docker [20] are designed to be an improvement in this regard, sacrificing the security and safety of complete operating system kernel separation for a greatly diminished footprint, much quicker launch times and lower management overhead [21].

The large-scale deployment of 'containerized' applications on a supercomputer is quite feasible, but for one notable caveat: the guest application is still using a limited subset of the interfaces exposed by the host operating system's kernel, and thus running a containerized Windows application in this manner would require a Windows host.

Luckily there exists a third option when it comes to launching a Windows environment on a Linux based supercomputer. The Wine project [22], [23] is an open source software compatibility layer designed to translate Windows semantics into their POSIX equivalents; from system calls to library APIs, file paths and named pipes to network and UNIX sockets, Wine seamlessly enables a myriad of Microsoft Windows applications to run on Linux and FreeBSD systems, in many cases with near-native performance and equivalent functionality. Additionally, because Wine is merely translating one software interface to another, there is very little in the way of environmental setup to perform prior to launching an application when compared to an equivalent task in a hardware virtual machine or with OS-assisted virtualization.

The goal of this approach is to present an unmodified Windows application with all of the appropriate software interfaces and a runtime environment that should be virtually indistinguishable from a real Windows operating system. A simplified depiction of Wine's layered architecture is presented in Figure 1.

### III. LLMAPREDUCE: MULTI-LEVEL MAP-REDUCE

The Wine environment provides a potentially efficient means for running Windows applications on a Linux-based supercomputer. Interactively launching many simultaneously Wine environments requires an effective means of coordinating the launch of thousands of these environments. Recent experiments have shown that the naive, serial job submission performance of a modern job scheduler can significantly slow down processing for jobs with a very large number of tasks [24]. To achieve maximal job launch performance for large HPC (high performance computing) or HPDA (high performance data analysis) jobs requires employing a technique known as multilevel scheduling which involves modifying our analysis code slightly to be able to process multiple datasets or files with a single job launch [25].

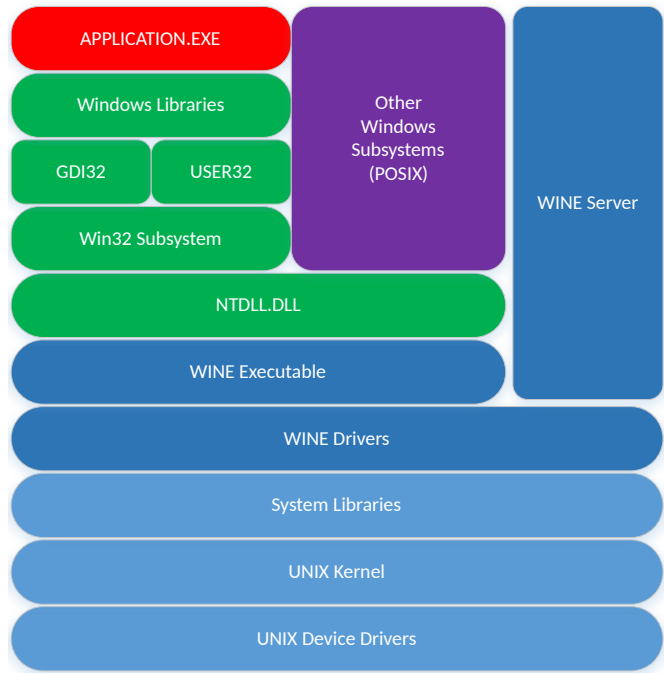

Fig. 1: Schematic depicting the architecture of a Windows program (APPLICATION.EXE) run on a UNIX-like system through the Wine emulator, diagram rendered from an ASCII version located in the Wine developer's manual. At the bottom of the figure is the native UNIX kernel, device drivers and system libraries of the host operating system. From there, the Wine UNIX binaries and user-space libraries are loaded, creating the virtual Windows environment which is then able to load Wine's implementation of the basic Win32 functionality provided by NTDLL.DLL, KERNEL.DLL, GDI.DLL and USER.DLL. This emulated Windows environment then loads the unmodified native Windows application used to invoke Wine and any Windows userspace libraries it requires. This environment is supported by the Wine server, depicted on the right, which provides inter-process communication, synchronization and process management.

The map-reduce parallel programming model [26] has become extremely popular in the big data community. The output of many workloads can increase greatly when run in an pleasingly parallel manner on a supercomputer. The LLMapReduce tool developed as part of the MIT SuperCloud software stack provides access to this familiar map-reduce programming model via a dramatically simplified interface and efficiently launches large, multi-level array jobs onto a cluster often reducing complex parallel scheduling, job submission and dependency resolution tasks into a single line of code while simultaneously maximizing job launch performance by reducing per-task latency [27]. Most importantly, LLMapReduce is not bound to a particular language and works with any executable, which makes LLMapReduce ideal for launching many simultaneous Wine instances. A notional illustration depicting the life cycle of the various components constituting a scheduler array job as generated by the LLMapReduce tool is presented in Figure 2.

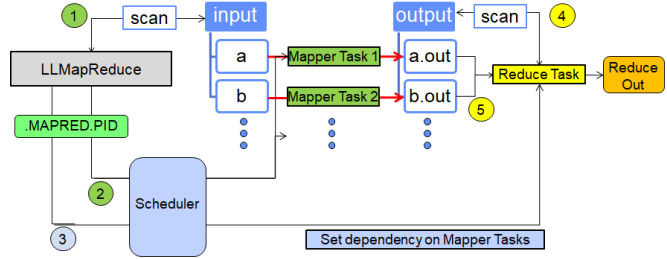

Fig. 2: Schematic depicting the life cycle of tasks launched using LLMapReduce (adapted from [27]). LLMapReduce scans an input directory and, for each file contained within, generates a job submission script within a scheduler array job and submits the aggregate batch job to the scheduler. Upon successful termination of all tasks, a "reduce" job is launched which can perform post-processing or epilog cleanup tasks.

Supercomputing systems require efficient mechanisms for managing the operational tasks involved in a program's life cycle on a large shared computing infrastructure: rapidly identifying available computing resources, allocating those resources to programs, scheduling the execution of those programs on their allocated resources, launching them, monitoring their execution and performing epilog clean-up tasks upon the program's termination (see Figure 3). The open source SLURM software provides these services and is independent of programming language (C, Fortran, Java, Matlab, etc.) or parallel programming model (message passing, distributed arrays, threads, map/reduce, etc.), which makes it ideal for launching Wine instances.

SLURM is an extremely scalable, full-featured Linux job scheduler with a modern, multi-threaded core scheduling engine and a very high-performance plug-in module architecture. [28] The combined feature set and serial launch latency of the SLURM scheduler compares favorably with other HPC resource managers [25], and it is well suited to managing a heterogeneous environment like the MIT SuperCloud.

## IV. EXPERIMENTAL ENVIRONMENT

The MIT Lincoln Laboratory Supercomputing Center provides (LLSC) a high-performance computing platform to over 1000 users at MIT, and is heavily focused on highly iterative interactive supercomputing and rapid prototyping workloads [29], [30]. A part of the LLSC mission to deliver new and innovative technologies and methods for enabling scientists and engineers to quickly ramp up the pace of their research. By leveraging supercomputing and big data storage assets the LLSC has built the MIT SuperCloud, a coherent fusion of the four largest computing ecosystems: supercomputing, enterprise computing, big data, and traditional databases. The MIT SuperCloud has spurred the development of a number of cross-ecosystem innovations in high performance databases [31], [32], database management [33], data protection [34], database federation [35], [36], data analytics [37] and system monitoring [38].

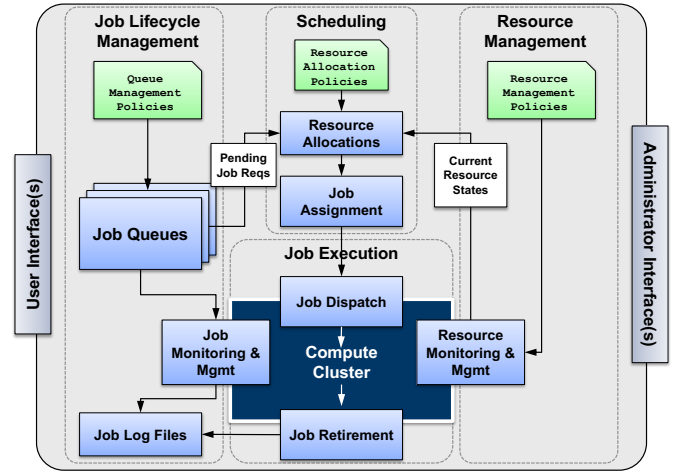

Fig. 3: Schematic depicting key components of a canonical cluster scheduler including job lifecycle management, resource management, task scheduling and job execution (adapted from [25]). The SLURM scheduler used on the MIT SuperCloud systems behaves according to this model.

All the experiments described in this paper were performed on the LLSC TX-Green Supercomputer using the MIT SuperCloud environment. The TX-Green supercomputer is a petascale system that consists of a heterogeneous mix of AMD Opterton, Intel Xeon, Nvidia, and Intel Xeon Phi processors connected to a single, non-blocking 10 Gigabit Ethernet Arista DCS-7508 core switch. All of the compute nodes used to launch the Windows instances were Intel Xeon Phi 7210 (Knight's Landing) processors with 64 cores, 192 GB of system RAM, 16 GB of on-package MCDRAM configured in 'flat' mode, and 4 TB of local storage. The Lustre [39] central storage system uses a 10 petabyte Seagate ClusterStor CS9000 storage array that is directly connected to the core switch, as is each individual cluster node. This architecture provides high bandwidth to all the nodes and the central storage, and is depicted in Figure 4.

## V. PERFORMANCE RESULTS

Rapid launching of Windows instances is prerequisite for running Windows applications interactively on a supercomputer. The launch times of the Wine environment on the MIT SuperCloud were obtained by running on a supercomputer consisting of 648 compute nodes, each with at least 64 Xeon Phi processing cores, for a total of 41,472 processing cores. In all cases, a single Windows instance was run on 1, 2, 4,..., and 256 compute nodes, followed by running 2, 4, ..., and 64 Windows instances on each of the 256 compute nodes to achieve a maximum of 16,384 simultaneous instances. An essential step for enabling rapid interactive launch is copying a typically several MB Windows executable and it's supporting environment (e.g. libraries and configuration files) from the user's home directory on the central storage to the local storage on each node. The copy time for this operation is shown in Figure 5 and is small compared to the launch time. This short copy time is achievable because the central file system is capable of a large amount of parallel I/O to the compute nodes. This parallel I/O rate is attainable when the copy is initiated from each of the target compute nodes and thus requires coordination within the overall parallel execution.

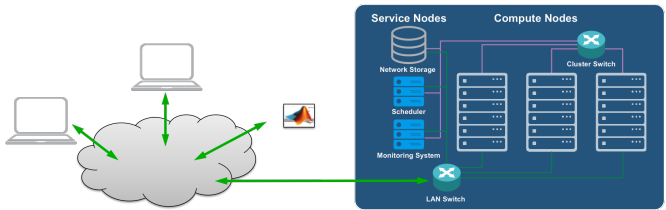

Fig. 4: Architecture of the MIT SuperCloud system. Users connect to the system over either a local area network or a wide area network. At the time of connection, their system joins the MIT SuperCloud and can act as a compute node in order to run parallel programs interactively. The centerpiece of the MIT SuperCloud is several file systems (Seagate, DDN, Dell, Hadoop, and Amazon S3) running on several different network fabrics (10 GigE, InfiniBand, OmniPath).

The launch times and launch rates of the Windows instances are shown in Figures 5, 6 and 7 along with data taken from the literature for launching for Windows instances on Azure [12] and Linux VM instances using Eucalyptus [14]. These results show that high launch Windows instance launch rates are achievable using Wine with LLMapReduce on the MIT SupercCloud. Launching over 16,000 instances in approximately 5 minutes directly enables interactive simulation, machine learning, and data analysis applications that require Windows executables.

## VI. SUMMARY AND FUTURE WORK

Traditional methods of scaling Microsoft Windows applications to run on thousands of processors have typically relied on heavyweight platform virtualization software that can be inefficient and slow to launch on modern manycore processors. Simulation, machine learning, and data analysis require a wide range of software which often depends upon specific operating systems, such as Microsoft Windows. Running this software interactively on massively parallel supercomputers can present many challenges. This paper describes a unique approach using the Lincoln Laboratory LLMapReduce technology in combination with the Wine Windows compatibility layer to rapidly launch and run Microsoft Windows applications on thousands of cores on a supercomputer. Specifically, this work demonstrates launching 16,000 Microsoft Windows applications in 5 minutes running on 16,000 processor cores. This capability significantly broadens the range of applications that can be run at large scale on a supercomputer. Future work will focus on extending this capability to larger numbers of cores running more diverse applications.

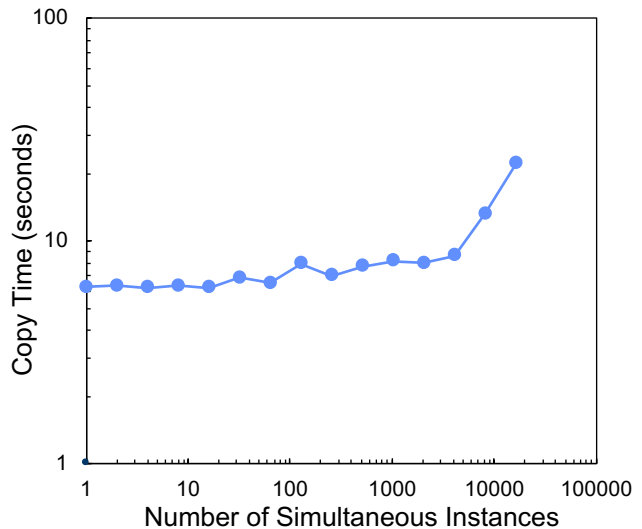

Fig. 5: Copy time of the Windows application from central storage to the local storage on each compute node versus the number of simultaneous instances launched.

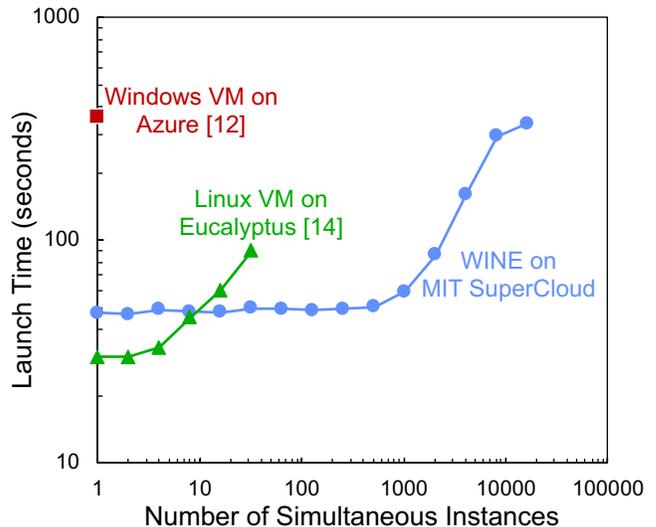

Fig. 6: Launch time of Wine on the MIT SuperCloud versus the number of simultaneous instances launched. Launch times for Windows instances on Azure [12] and Linux VM instances using Eucalyptus [14] are also shown.

## ACKNOWLEDGMENTS

The authors wish to acknowledge the following individuals for their contributions: Bob Bond, Alan Edelman, Jack Fleischman, Charles Leiserson, Dave Martinez, and Paul Monticciolo.

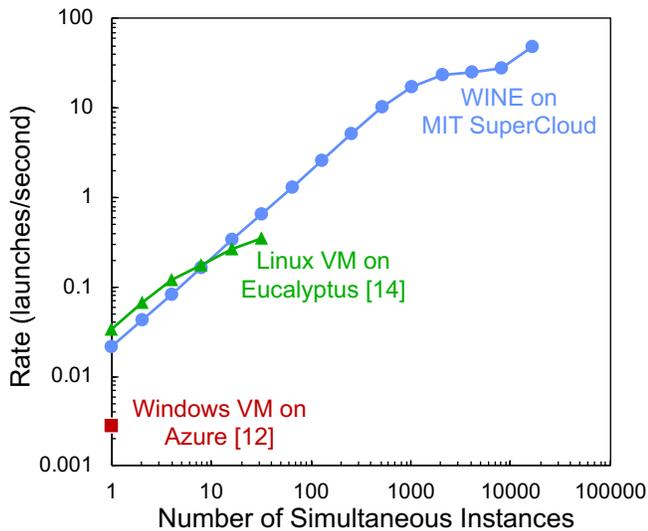

Fig. 7: Launch rate of Wine on the MIT SuperCloud versus the number of simultaneous instances launched. Launch rates for Windows instances on Azure [12] and Linux VM instances using Eucalyptus [14] are also shown.